# Magnetic-polarization-dependent spectroscopy of lanthanide-doped anisotropic crystals

ZOE LIESTMANN, LUCA KOLDEWEYH, MORITZ BADTKE†, SASCHA KALUSNIAK, CHRISTIAN KRÄNKEL, AND HIROKI TANAKA*

*Leibniz-Institut für Kristallzüchtung (IKZ), Max-Born-Str. 2, 12489 Berlin, Germany*
*†current affiliation: Fraunhofer Institute of Optronics, System Technologies and Image Exploitation (IOSB), Gutleuthausstr. 1, 76275 Ettlingen, Germany*
**hiroki.tanaka@ikz-berlin.de*

**Abstract:** We experimentally demonstrate that absorption and emission spectra of trivalent lanthanide-doped anisotropic crystals can exhibit a significant magnetic-polarization dependence, which has been largely overlooked in spectroscopic studies to date. Focusing on the uniaxial laser host $LiYF_4$ (YLF) doped with $Yb^{3+}$, $Tm^{3+}$, $Er^{3+}$, and $Ho^{3+}$, we measure magnetic-polarization-dependent absorption and emission spectra for transitions with strong magnetic-dipole (MD) contributions predicted by theory. Our results reveal that MD-induced spectral anisotropy, *i.e.*, spectral differences for the same electric field orientation but for different magnetic field orientations, is present even in these well-established laser materials. A complete spectroscopic characterization of uniaxial crystals requires three polarizations, including the α-polarization, with both the electric field vector **E** and the magnetic field vector **H** perpendicular to the *c*-axis ($\mathbf{E} \perp c$, $\mathbf{H} \perp c$), in addition to the commonly used two polarizations π ($\mathbf{E} \parallel c$, $\mathbf{H} \perp c$) and σ ($\mathbf{E} \perp c$, $\mathbf{H} \parallel c$). We further discuss the observed MD-induced spectral anisotropy and calculated MD branching ratios, the impact of the anisotropy on emission cross-section calculations, and the relevance of our results to other uniaxial and biaxial crystals.

## 1. Introduction

Crystalline materials doped with trivalent lanthanide (rare-earth) ions have been extensively studied over the past decades owing to their optical properties originating from intra-configurational 4*f*-4*f* transitions. Lanthanide-doped crystals play a central role in a wide range of applications in optics and photonics, including solid-state lasers [1,2], fluorescence phosphors [3,4], scintillators [5], and solid-state laser cooling [6]. Accurate knowledge of absorption and emission cross-section spectra is essential for all of these applications. Particularly for gain media in solid-state lasers, these material parameters determine pump absorption efficiency, optical gain, accessible laser wavelengths, tunability, and the achievable pulse duration in mode-locked operation.

In optically anisotropic (uniaxial or biaxial) host crystals, the optical spectra depend on the polarization of the absorbed or emitted light. In this context, polarization refers to the orientation of the *electric* field vector relative to the crystals' optical axes. Hereafter, this is explicitly denoted as *electric polarization*. Observed absorption and emission associated with 4*f* electrons are predominantly caused by electric-dipole (ED) transitions, which are in most cases orders of magnitude stronger than magnetic-dipole (MD) transitions and therefore dominate the spectra of most lanthanide-doped materials. This often justifies the use of electric-polarization-dependent spectra.

On the other hand, certain 4*f*-4*f* transitions are known to exhibit substantial MD contributions [7,8]. In Judd-Ofelt analyses [9,10], accounting for MD transitions is often essential; otherwise, the ED line strengths of transitions with significant MD contributions may be overestimated, affecting the calculated Judd-Ofelt parameters. Strong MD contributions were experimentally identified, *e.g.*, in the cubic crystal $Y_2O_3$ doped with $Eu^{3+}$, $Dy^{3+}$, or $Tm^{3+}$ [11,12].

However, the influence of MD transitions on spectral anisotropy has rarely been considered. MD contributions make the absorption and emission spectra depend not only on the electric polarization but also on the *magnetic polarization* in anisotropic crystals. This results in three and six polarization-dependent spectra for uniaxial and biaxial crystals, respectively. This aspect has so far been largely overlooked because the contribution of MD transitions is often negligible. A notable exception is $Eu^{3+}$, which shows a strong MD contribution in its visible emission, as several corresponding transitions are ED-forbidden owing to the selection rules for the emitting level with a total angular momentum quantum number $J = 0$ (see Tab. 1) [13]. In addition, Chen identified magnetic-polarization dependence in the visible absorption and emission spectra of $Tb^{3+}$ and $Dy^{3+}$ in various oxide crystals [14]. Near-infrared transitions of $Er^{3+}$ and $Ho^{3+}$ are also known to exhibit considerable MD contributions. Therefore, Payne *et al.* suggested the need for magnetic-polarization-dependent spectroscopy [15]. A strong magnetic-polarization dependence was recently also found in the biaxial $Yb^{3+}$:$LaLuO_3$, where six distinct polarization-dependent spectra were identified [16]. Its strong MD contribution was attributed to the inversion symmetry of the $Lu^{3+}$ sites, where parts of the $Yb^{3+}$ ions are incorporated; the corresponding transitions remain ED-forbidden by the Laporte rule.

Here, we demonstrate that the magnetic-polarization dependence is present across a wide range of transitions in anisotropic crystals doped with trivalent lanthanide ions. We investigate the MD-induced spectral anisotropy, *i.e.*, magnetic-polarization dependence under fixed electric polarization, in the well-known uniaxial laser host crystal $LiYF_4$ (YLF) doped with $Yb^{3+}$, $Tm^{3+}$, $Er^{3+}$, or $Ho^{3+}$. These crystals are widely used as gain media for near- and mid-infrared solid-state lasers [17]. Contrary to the common assumption that only two polarized spectra exist, we found three distinct polarization-dependent spectra in these uniaxial crystals caused by MD contributions.

In this work, we first revisit the theory of MD transition probabilities and selection rules. Based on the theoretical calculations, we identify transitions with high MD contributions that are expected to exhibit MD-induced spectral anisotropy. By examining four transitions in absorption and seven in emission in the four lanthanide ions, we experimentally confirm that MD-induced anisotropy is significant for all selected transitions. We further discuss the correlation between the theoretical MD branching ratio and observed MD-induced spectral anisotropy, the impact on calculated emission cross-sections, the relevance of magnetic-polarization-dependent spectroscopy to other host crystals, and implications for solid-state lasers.

## 2. Magnetic-dipole contribution in 4*f*-4*f* transitions

According to the Laporte selection rule [18], intra-configurational ED transitions within the 4*f* shell are forbidden in free ions, since ED transitions can only occur between states with opposite parity. However, in solids with a non-centrosymmetric crystal field, parity mixing of 4*f* and 5*d* states allows 4*f*-4*f* transitions with peak transition cross-sections typically on the order of $10^{-22}$–$10^{-18}$ $cm^2$. In contrast, MD transitions can occur between states of the same parity; thus, 4*f*-4*f* transitions are allowed for free ions, but they typically exhibit smaller peak cross-sections of around $10^{-24}$–$10^{-22}$ $cm^2$. Table 1 summarizes the selection rules for ED and MD transitions, derived from the Wigner-Eckart theorem, which are typically used in the Judd–Ofelt theory [19]. According to these rules, transitions between manifolds with $\Delta J = \pm 1$ within the same $LS$ term ($\Delta L = 0$) are MD-allowed. This is the case for the ground-state absorption transition to the lowest excited state in most trivalent lanthanides except for $Gd^{3+}$ and $Tm^{3+}$. Note that electric-quadrupole transitions can also occur between states of the same parity; however, their transition rates are significantly lower than those of MD transitions [8].

**Table 1. Selection rules for electric-dipole and magnetic-dipole transitions used in Judd–Ofelt theory** [19].

| | $S$ | $L$ | $J$ (0 ↔ 0 is forbidden) | Parity |
|---|---|---|---|---|
| Electric dipole | $\Delta S = 0$ | $\Delta L \le 6$ | $\Delta J \le 6$ | opposite |
| Magnetic dipole | $\Delta S = 0$ | $\Delta L = 0$ | $\Delta J = 0, \pm 1$ | same |

Oscillator strengths $f'_{\mathrm{MD}}$ and spontaneous emission rates $A'_{\mathrm{MD}}$ of MD transitions for free ions are given by:

$$f'_{\mathrm{MD}} = \frac{8\pi^2 m_e c}{3h\lambda(2J+1)} S_{\mathrm{MD}}, \tag{1}$$

$$A'_{\mathrm{MD}} = \frac{64\pi^4 e^2}{3h\lambda^3(2J'+1)} S_{\mathrm{MD}}, \tag{2}$$

where $h$ is the Planck constant, $c$ the speed of light, $e$ the electron charge, $m_e$ the electron mass, $\lambda$ the wavelength of the transition, and $J$ and $J'$ the total angular momentum quantum number of the upper and lower manifold, respectively. $S_{\mathrm{MD}}$ is the MD line strength given by [19]:

$$S_{\mathrm{MD}} = \left(\frac{h}{4\pi m_e c}\right)^2 |\langle [SL]J \| L + g_e S \| J'[S'L'] \rangle|^2. \tag{3}$$

Here, $\langle [SL]J \| L + g_e S \| J'[S'L'] \rangle$ is the reduced matrix element of the MD operator between intermediate-coupled states, where $g_e$ is the g-factor of the electron, approximately equal to 2.0. Note that Eqs. (1)-(3) are written in CGS units, as in [19]. In a solid with refractive index $n$, the MD oscillator strength $f_{\mathrm{MD}}$ and spontaneous emission rate $A_{\mathrm{MD}}$ scale as $f_{\mathrm{MD}} = n^3 f'_{\mathrm{MD}}$ and $A_{\mathrm{MD}} = n^3 A'_{\mathrm{MD}}$, respectively [20]. Carnall *et al.* tabulated the MD oscillator strengths of selected ground-state absorption transitions for free lanthanide ions [7]. Dodson *et al.* further extended this table and also tabulated the spontaneous MD emission rates for transitions in the wavelength range between 300 nm and 1700 nm [8].

Figure 1 shows the energy level diagrams of $Yb^{3+}$, $Tm^{3+}$, $Er^{3+}$, and $Ho^{3+}$. The absorption and emission transitions selected in this work to investigate the MD-induced spectral anisotropy are indicated with up- and downward arrows, respectively. These selected transitions fulfill the MD selection rule in Table 1 and are expected to show comparably strong MD contributions. We excluded the $^3H_5 \rightarrow {}^3H_6$ emission transition of $Tm^{3+}$ at ~1200 nm because of its low emission intensity due to multi-phonon relaxation to the $^3F_4$ manifold.

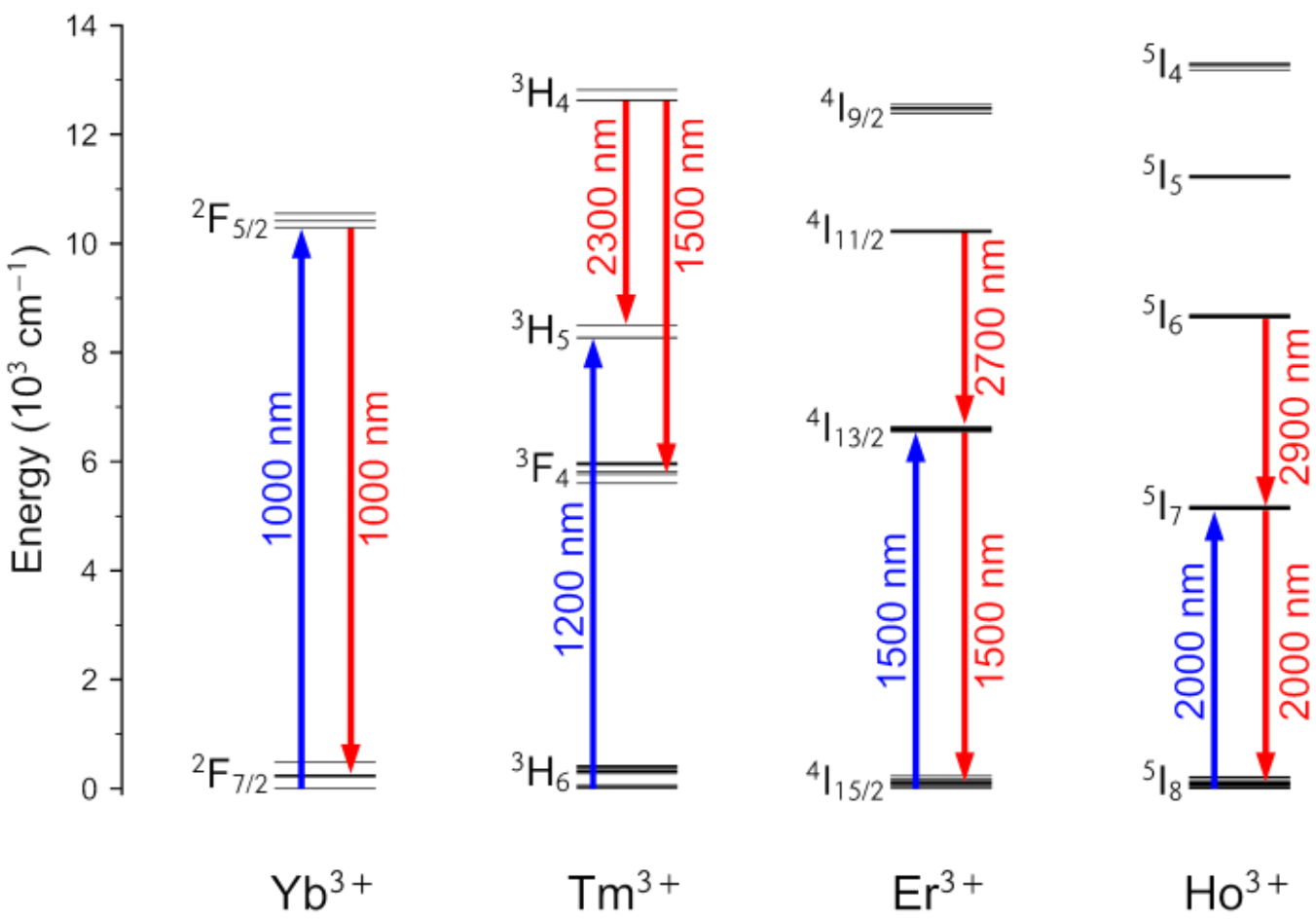

Fig. 1: Energy level diagrams of $Yb^{3+}$, $Tm^{3+}$, $Er^{3+}$, and $Ho^{3+}$ with the transitions investigated in this work indicated by arrows.

Note that the energy manifolds shown in Fig. 1 are labeled according to the conventional *LS*-coupling, although intermediate coupling (considering *jj*-coupling) may be required for a more accurate description. *LS*-coupling is accurate for $Yb^{3+}$, and provides a good approximation for the energy manifolds of $Er^{3+}$ and $Ho^{3+}$ relevant to this work [21,22]. In contrast, $Tm^{3+}$ exhibits significant mixing of *LS*-coupled states with $J = 4$, *e.g.*, for a free $Tm^{3+}$ ion, the $^3H_4$ state under intermediate-coupling comprises 27% of the *LS*-coupled $^3F_4$ state [23]. This mixing leads to a strong MD contribution in the transition $^3H_4 \rightarrow {}^3F_4$ around 1500 nm, although this transition is not MD-allowed under pure *LS*-coupling.

Table 2 summarizes the mean absorption wavelengths $\bar{\lambda}_{abs}$ and calculated MD oscillator strengths of the ground-state absorption transitions into the first excited state in free $Yb^{3+}$, $Er^{3+}$, $Tm^{3+}$, and $Ho^{3+}$ ions. Table 3 summarizes the mean emission wavelengths $\bar{\lambda}_{em}$, the spontaneous emission rates of the transition $A(JJ')$, given by $\beta(JJ')/\tau_{rad}$ (where $\tau_{rad}$ is the radiative lifetime of the emitting manifold and $\beta(JJ')$ is the branching ratio for the transition), as well as the spontaneous MD emission rates $A_{MD}$ for the seven emission transitions investigated in this work. The spontaneous MD emission rates $A_{MD}$ are taken from the literature where available; otherwise, they are calculated using Eqs. (2) and (3). We calculated the MD branching ratio $\beta_{MD}$, *i.e.*, the fraction of MD transition given by $A_{MD}/A(JJ')$, where the total spontaneous emission rate is $A(JJ') = A_{ED} + A_{MD}$.

For transitions with a branching ratio $\beta(JJ')$ of 100%, we utilized experimentally determined radiative lifetimes. For the $^5I_7$ manifold of Ho:YLF, we used our measured radiative lifetime of 15.2 ms. Note that some literature data for the radiative lifetimes and branching ratios may require reassessment, particularly for the 2.7-µm transition in $Er^{3+}$ and the 2.9-µm transition in $Ho^{3+}$, because of the unknown non-radiative decay rates. For these transitions, a large discrepancy was identified between experimentally determined values and those from the Judd–Ofelt analysis [24]. For the $^3H_4$ manifold of $Tm^{3+}$, we adopted a radiative lifetime of 2.0 ms, determined for a Tm(0.5%):YLF [25], instead of the value of 1.44 ms from the Judd–Ofelt analysis [26]. The branching ratios of the manifold were taken from [26].

**Table 2. Mean absorption transition wavelengths $\bar{\lambda}_{abs}$ and free-ion magnetic-dipole (MD) oscillator strengths $f'_{MD}$ for ground-state absorption transitions of free $Yb^{3+}$, $Tm^{3+}$, $Er^{3+}$, and $Ho^{3+}$ ions with MD oscillator strengths larger than $10^{-7}$. The values of $f'_{MD}$ are taken from [8].**

| Ion | Lower manifold *SLJ* | Upper manifold *S'L'J'* | $\bar{\lambda}_{abs}$ (nm) | $f'_{MD}$ ($\times 10^{-8}$) [8] |
|---|---|---|---|---|
| $Yb^{3+}$ | $^2F_{7/2}$ | $^2F_{5/2}$ | 961 | 17.76 |
| $Tm^{3+}$ | $^3H_6$ | $^3H_5$ | 1200 | 27.25 |
| $Er^{3+}$ | $^4I_{15/2}$ | $^4I_{13/2}$ | 1515 | 30.82 |
| $Ho^{3+}$ | $^5I_8$ | $^5I_7$ | 1968 | 29.47 |

**Table 3. Mean emission transition wavelength $\bar{\lambda}_{em}$, room-temperature radiative lifetime $\tau_{rad}$ of the emitting manifold, branching ratio $\beta(JJ')$, spontaneous emission rate $A(JJ')$, magnetic-dipole (MD) spontaneous emission rate $A_{MD}$, and calculated MD branching ratio $\beta_{MD}$ for the seven investigated emission transitions of $Yb^{3+}$, $Tm^{3+}$, $Er^{3+}$, and $Ho^{3+}$ in YLF. For calculating $A_{MD}$, we used the polarization-averaged refractive index $\bar{n}$ based on the reported Sellmeier equation [27]. For wavelengths above 2600 nm, the refractive indices were extrapolated using the equation.**

| Ion | Lower manifold *SLJ* | Upper manifold *S'L'J'* | $\bar{\lambda}_{em}$ (nm) | $\tau_{rad}$ (ms) | $\beta(JJ')$ (%) | $A(JJ')$ ($s^{-1}$) | $A_{MD}$ ($s^{-1}$) | $\beta_{MD}$ (%) |
|---|---|---|---|---|---|---|---|---|
| $Yb^{3+}$ | $^2F_{7/2}$ | $^2F_{5/2}$ | 999 | 2.2 [28] | 100 | 454 | 49.5 [8] | 11 |
| $Tm^{3+}$ | $^3F_4$ | $^3H_4$ | 1468 | 1.9 [25] | 10.2 [29] | 53.7 | 36.3 [29] | 67 |
| | $^3H_5$ | $^3H_4$ | 2325 | 1.9 [25] | 2.4 [26] | 11.7 | 5.4 [29] | 46 |
| $Er^{3+}$ | $^4I_{15/2}$ | $^4I_{13/2}$ | 1548 | 10 [30] | 100 | 100 | 30.4 [8] | 30 |
| | $^4I_{13/2}$ | $^4I_{11/2}$ | 2741 | 9.8 [31] | 38.7 [32] | 39.5 | 20.3 | 51 |
| $Ho^{3+}$ | $^5I_8$ | $^5I_7$ | 2009 | 15.2 | 100 | 66 | 47.5 [29] | 72 |
| | $^5I_7$ | $^5I_6$ | 2889 | 6.4 [29] | 13.5 [29] | 21 | 18.6 | 88 |

ED- and MD-transitions between Stark sub-levels in different energy manifolds obey selection rules associated with their full-rotational symmetries. In YLF, lanthanide ions occupy $S_4$ point-group symmetry sites, and the manifolds split according to this symmetry. Manifolds with integer total angular momentum quantum number *J*, *i.e.*, in $Tm^{3+}$ and $Ho^{3+}$, split into Stark sub-levels transforming as $\Gamma_1$ (= $A_1$), $\Gamma_2$ (= $A_2$), and $\Gamma_{3,4}$ (= E). For half-integer *J*, *i.e.*, for $Yb^{3+}$ and $Er^{3+}$, the corresponding Kramers-degenerated Stark sub-levels are described by the double-group irreducible representations $\Gamma_{5,6}$ (= $E_{1/2}$) and $\Gamma_{7,8}$ (= $E_{2/3}$) in Koster (Mulliken) notation [33]. Tables 4 and 5 summarize the selection rules for integer and half-integer *J* under $S_4$ point-group symmetry, with **E** being the electric field vector and **H** the magnetic field vector. Table 4 is taken from [34], and Table 5 was derived using the multiplication table given in [33].

**Table 4. Selection rules for electric- and magnetic-dipole transitions under $S_4$ point group symmetry for integer *J*.**

| | $\Gamma_1$ (=$A_1$) | $\Gamma_2$ (=$A_2$) | $\Gamma_{3,4}$ (=E) |
|---|---|---|---|
| $\Gamma_1$ | **H** ‖ *c* | **E** ‖ *c* | **E** ⊥ *c*, **H** ⊥ *c* |
| $\Gamma_2$ | | **H** ‖ *c* | **E** ⊥ *c*, **H** ⊥ *c* |
| $\Gamma_{3,4}$ | | | **E** ‖ *c*, **H** ‖ *c* |

**Table 5. Selection rules for electric- and magnetic-dipole transitions under $S_4$ point group symmetry for half-integer *J*.**

| | $\Gamma_{5,6}$ (=$E_{1/2}$) | $\Gamma_{7,8}$ (=$E_{3/2}$) |
|---|---|---|
| $\Gamma_{5,6}$ | **E** ⊥ *c*, **H** ‖ *c*, **H** ⊥ *c* | **E** ‖ *c*, **E** ⊥ *c*, **H** ⊥ *c* |
| $\Gamma_{7,8}$ | | **E** ⊥ *c*, **H** ‖ *c*, **H** ⊥ *c* |

These tables indicate that certain transitions between Stark sub-levels show magnetic-polarization dependence. For integer *J* (Table 4), transitions between levels with corresponding representations $\Gamma_1 \leftrightarrow \Gamma_1$, $\Gamma_1 \leftrightarrow \Gamma_{3,4}$, $\Gamma_2 \leftrightarrow \Gamma_2$, $\Gamma_2 \leftrightarrow \Gamma_{3,4}$, and $\Gamma_{3,4} \leftrightarrow \Gamma_{3,4}$ may exhibit MD-induced spectral anisotropy because the selection rules are fulfilled only in one of the magnetic polarizations, **H** ‖ *c* or **H** ⊥ *c*. For half-integer *J* (Table 5), the same holds for transitions between levels with representations $\Gamma_{5,6} \leftrightarrow \Gamma_{7,8}$. In contrast, transitions $\Gamma_{5,6} \leftrightarrow \Gamma_{5,6}$ and $\Gamma_{7,8} \leftrightarrow \Gamma_{7,8}$ fulfill the MD selection rule for both magnetic polarizations, and thus significant MD-induced spectral anisotropy is not expected. However, since the selection rules do not provide quantitative transition strengths, spectral anisotropy may still appear in these cases.

## 3. Experimental methods

### *3.1 Crystal growth and sample preparation*

We grew YLF crystals doped with $Yb^{3+}$, $Tm^{3+}$, $Er^{3+}$, and $Ho^{3+}$ by the Czochralski method. As the starting materials, the following compounds with respective purities were used: LiF crystals (Korth Kristalle GmbH), $YF_3$ powder (5N, AC Materials, Inc. or Projector GmbH), $YbF_3$ powder (5N, AC Materials, Inc.), $TmF_3$ powder (5N, Projector GmbH), $ErF_3$ powder (4N, Johnson Matthey), $HoF_3$ powder (5N, Projector GmbH). To prevent the peritectic crystallization of $YF_3$, the starting materials were slightly enriched with LiF relative to the stoichiometry [35]. The doping concentrations were 5 at.% for Yb:YLF and Er:YLF, and 1 at.% for Tm:YLF and Ho:YLF. The segregation coefficients were confirmed to be near unity for all doping ions.

The grown crystals were oriented along their crystallographic axes using Laue diffraction patterns, and then cut and polished to prepare *a*-cut and *c*-cut samples for spectroscopy. In the presence of significant MD contribution, the full spectroscopic characterization of a uniaxial crystal requires measurements for three polarizations: The conventional polarizations π ($\mathbf{E} \parallel c$, $\mathbf{H} \perp c$) and σ ($\mathbf{E} \parallel c$, $\mathbf{H} \perp c$), as well as ($\mathbf{E} \perp c$, $\mathbf{H} \perp c$), hereafter denoted as α-polarization [13]. *a*-cut samples provide access to the π- and σ-polarized spectra, whereas *c*-cut samples provide access only to the α-polarized spectra. Figure 2 visualizes these three polarizations in a tetragonal structure, with ***k*** being a wave vector.

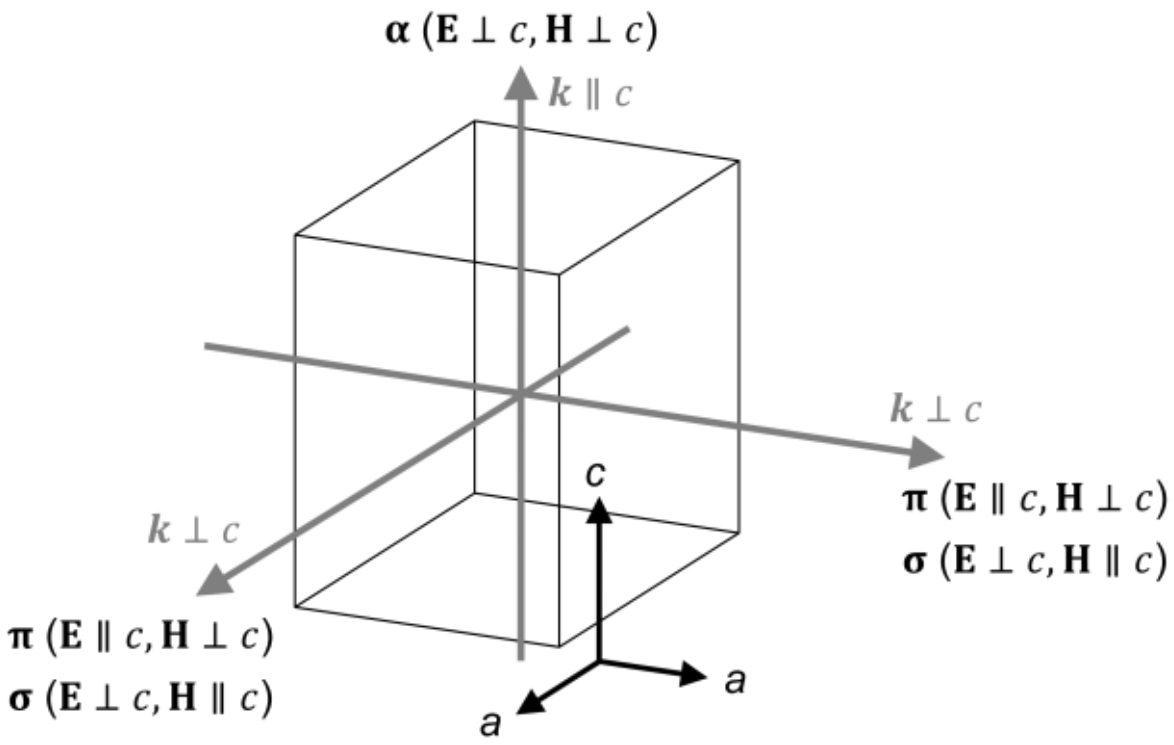

Fig. 2. Schematic illustration of the three polarizations in a tetragonal structure.

### *3.2 Absorption spectroscopy*

For the ground-state absorption transitions listed in Table 2, the three polarized transmission spectra were measured using a double-beam spectrophotometer (PerkinElmer, Lambda1050) equipped with Glan-Thompson polarizers. After correcting the measured transmittance for the Fresnel reflections using the refractive indices of YLF [27], we calculated the absorption coefficient, $\alpha_{abs}$, using the Beer-Lambert law. Absorption cross-sections $\sigma_{abs}$ were then calculated from the absorption coefficient and the density of the doped lanthanide ion, $N$, according to $\sigma_{abs} = \alpha_{abs}/N$.

### *3.3 Fluorescence spectroscopy*

We recorded three polarized fluorescence spectra for each crystal using *a*- and *c*-cut samples. Yb:YLF, Tm:YLF, and Er:YLF were excited at 895 nm, 793 nm, and 972 nm, respectively, using a continuous-wave Ti:sapphire laser (MSquared Lasers, SolsTiS), whereas Ho:YLF was excited using a blue laser diode (Nichia, NDB7K75) at a wavelength of 453 nm. The emitted fluorescence was collected and imaged through a polarizer onto the entrance slit of a 1-m monochromator (Horiba, 1000M Series II) using a $CaF_2$ lens. In the monochromator, a near-infrared grating (1200 grooves/mm, blazed at 750 nm) was used for the 1-μm transition of Yb:YLF, whereas a mid-infrared grating (300 grooves/mm, blazed at 2000 nm) was used for all other fluorescence transitions investigated in this work. As detectors, we used a photomultiplier tube (PMT, Hamamatsu, R5108) for wavelengths up to 1200 nm, a thermoelectrically cooled InGaAs photodiode (Hamamatsu, G12183-220K) with a transimpedance pre-amplifier (Hamamatsu, C4159-03) for the range of 1300–2550 nm, and a liquid-nitrogen-cooled InSb detector (Horiba, DSS-IS020L) for wavelengths above 2550 nm.

The detection signal was read out through a digital lock-in amplifier (Stanford Research Systems, SR830) synchronized with the modulation of either the fluorescence signal or the excitation beam and calibrated by the spectral response function of the setup. For each sample, the spectral resolution was adjusted to fully resolve all spectral features.

To reduce atmospheric water vapor absorption, we minimized the distance between the sample and the monochromator and purged the monochromator with $N_2$ gas. To minimize the influence of reabsorption on the detected fluorescence, we used samples as thin as 200 µm for all transitions into the ground state.

Emission cross-section spectra were calculated from the fluorescence spectra using the Füchtbauer–Ladenburg (F–L) formula [36] for uniaxial crystals considering three polarized spectra:

$$\sigma_{\mathrm{em},\zeta}(\lambda) = \frac{\beta(JJ')\lambda^5}{8\pi c n_\zeta^2(\lambda)\tau_{\mathrm{rad}}} \frac{3I_\zeta(\lambda)}{\int \sum_{\zeta'} \lambda' I_{\zeta'}(\lambda') d\lambda'}, \tag{4}$$

where $I_\zeta(\lambda)$ is the fluorescence intensity spectrum for polarization $\zeta$ (= π, σ, α) and $n_\zeta$ is the refractive index. Where MD contributions are negligible, the relation $I_\sigma(\lambda) = I_\alpha(\lambda)$ can be used. All emission cross-section spectra presented in this work were calculated by Eq. (4) using the radiative lifetimes $\tau_{\mathrm{rad}}$ and branching ratios $\beta(JJ')$ listed in Table 3.

To accurately determine emission cross-sections, the relative fluorescence intensities in all three polarizations must be known. However, the α-polarized spectrum can only be measured using *c*-cut samples, whereas π- and σ-polarized spectra are measured using *a*-cut samples. Therefore, we applied one of the following procedures to accurately scale the intensities of fluorescence measured using two different samples.

For transitions terminating in the ground-state manifold, we first calculated the emission cross-section spectra with arbitrary relative fluorescence intensities, yielding cross sections with arbitrary scaling. We then determined their absolute scaling by comparison with those calculated from absorption cross-section spectra using the reciprocity (McCumber) relation [37]:

$$\sigma_{\mathrm{em},\zeta}(\lambda) = \sigma_{\mathrm{abs},\zeta}(\lambda) \frac{Z_{\mathrm{l}}}{Z_{\mathrm{u}}} \exp\left(-\frac{hc/\lambda - E_{\mathrm{ZPL}}}{k_B T}\right), \tag{5}$$

where $k_{\mathrm{B}}$ is the Boltzmann constant, $T$ is the temperature, $E_{\mathrm{ZPL}}$ is the zero-phonon-line energy, and $Z_{\mathrm{l}}$ and $Z_{\mathrm{u}}$ are the Boltzmann partition functions of the lower and upper manifolds. Comparison of the spectral shapes derived using these two distinct approaches also allows us to assess whether the measured fluorescence is free from reabsorption, since reabsorption would cause depletion, a decreased signal intensity on the short-wavelength side, in the spectra calculated by the F–L formula.

For transitions not terminating in the ground-state manifold, where possible, we extended the measurement range to include a spectrally adjacent emission line with negligible MD contribution, thus yielding identical σ- and α-polarized spectra. This enabled relative intensity scaling between fluorescence spectra obtained using *a*- and *c*-cut samples. Where such lines were not accessible, we measured fluorescence spectra using *a*- and *c*-cut samples of identical thickness under identical conditions and with equal absorbed power. This was done using **E** ⊥ *c*-polarized excitation at wavelengths with equal absorption for σ- and α-polarization.

## 4. Results

### *4.1 Yb:YLF*

Figures 3(a) and 3(b) show the polarized absorption and emission cross-section spectra of Yb:YLF, respectively. The differential spectra between the $\sigma$- and $\alpha$-polarization ($\Delta\sigma_{abs}$ and $\Delta\sigma_{em}$) show that the MD contribution is significant for the $^2F_{7/2} \leftrightarrow {}^2F_{5/2}$ transitions. In both absorption and emission, differences in cross section as large as $0.12\times10^{-20}$ and $0.15\times10^{-20}$ cm$^2$ are found at a wavelength of ≈972 nm, corresponding to the zero-phonon line [28].

To quantify the anisotropy between the σ- and α-polarizations associated with the MD contribution, we introduce the MD-induced spectral anisotropy factor $\kappa_{MD}$ defined as the integrated anisotropy normalized to the integral of the polarization-averaged spectra:

$$\kappa_{\mathrm{MD}} = \frac{3\int|\sigma_\alpha(\lambda)-\sigma_\sigma(\lambda)|d\lambda}{\int[\sigma_\pi(\lambda)+\sigma_\sigma(\lambda)+\sigma_\alpha(\lambda)]d\lambda}, \quad (6)$$

where $\sigma_\pi$, $\sigma_\sigma$, and $\sigma_\alpha$ are the polarized absorption or emission cross-sections. For Yb:YLF, $\kappa_{MD}$ is 6% for absorption and 7% for emission.

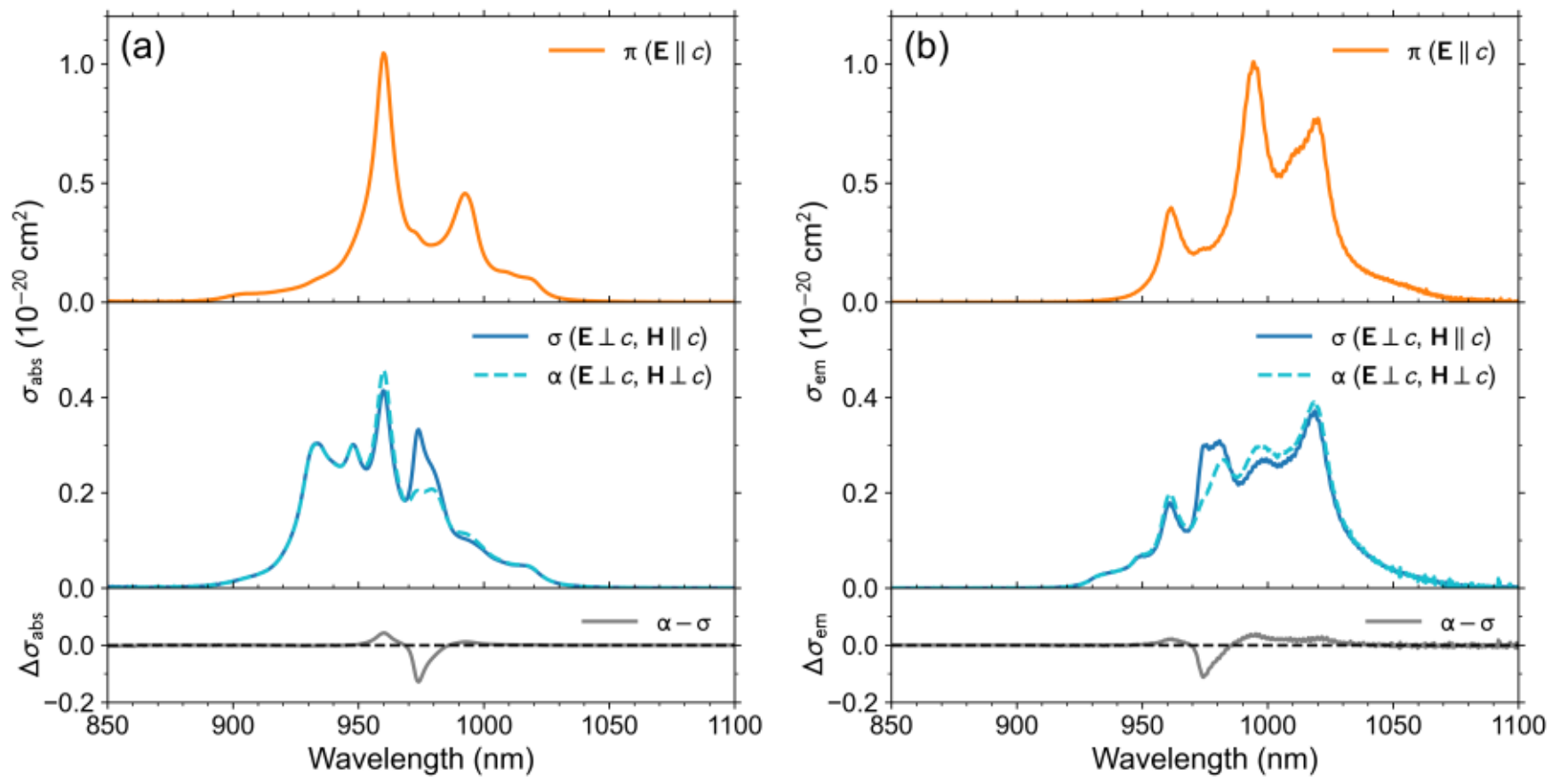

Fig. 3. Polarized absorption (a) and emission (b) cross-section spectra of Yb:YLF for the $^2F_{7/2} \leftrightarrow {}^2F_{5/2}$ transitions and corresponding differential cross-sections Δσ between the σ- and α-polarization.

### *4.2 Tm:YLF*

Figure 4 shows the polarized absorption cross-section spectra of Tm:YLF for the ground-state absorption transition $^3H_6 \rightarrow {}^3H_5$. Except for this transition, all σ- and α-polarized ground-state absorption transitions of Tm:YLF showed identical spectra over the measured wavelength range down to 250 nm, indicating negligible MD contributions. Note that no MD-induced spectral anisotropy was observed for the $^3H_6 \rightarrow {}^1I_6$ absorption transition, for which the free-ion MD oscillator strength $f'_{MD}$ is $1.4\times10^{-8}$ [8]. In contrast, the absorption peak at 1209 nm is nearly a factor of two higherfor σ-polarization than for α-polarization, indicating a strong MD contribution for the magnetic polarization $\mathbf{H} \parallel c$ compared with $\mathbf{H} \perp c$. The MD-induced spectral anisotropy factor $\kappa_{MD}$ is 16% for this absorption transition.

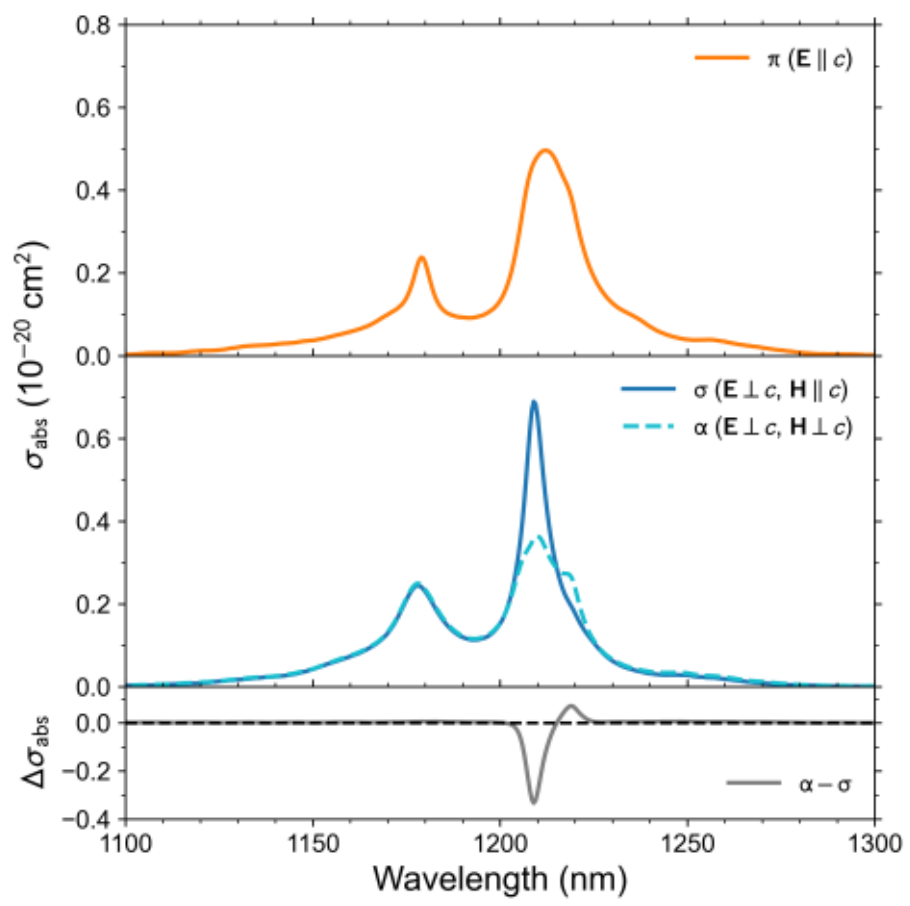

Fig. 4. Polarized absorption cross-section spectra of Tm:YLF for the ground-state absorption transition $^3H_6 \rightarrow {}^3H_5$ and corresponding differential cross-sections Δσ between the σ- and α-polarization.

Figure 5 shows the polarized emission cross-section spectra for the $^3H_4 \rightarrow {}^3F_4$ transition in Tm:YLF. A difference between σ- and α-polarizations of up to $0.15\times10^{-21}$ cm$^2$ is found at 1465 nm. For this transition, the overall cross-section is higher for the α-polarization.

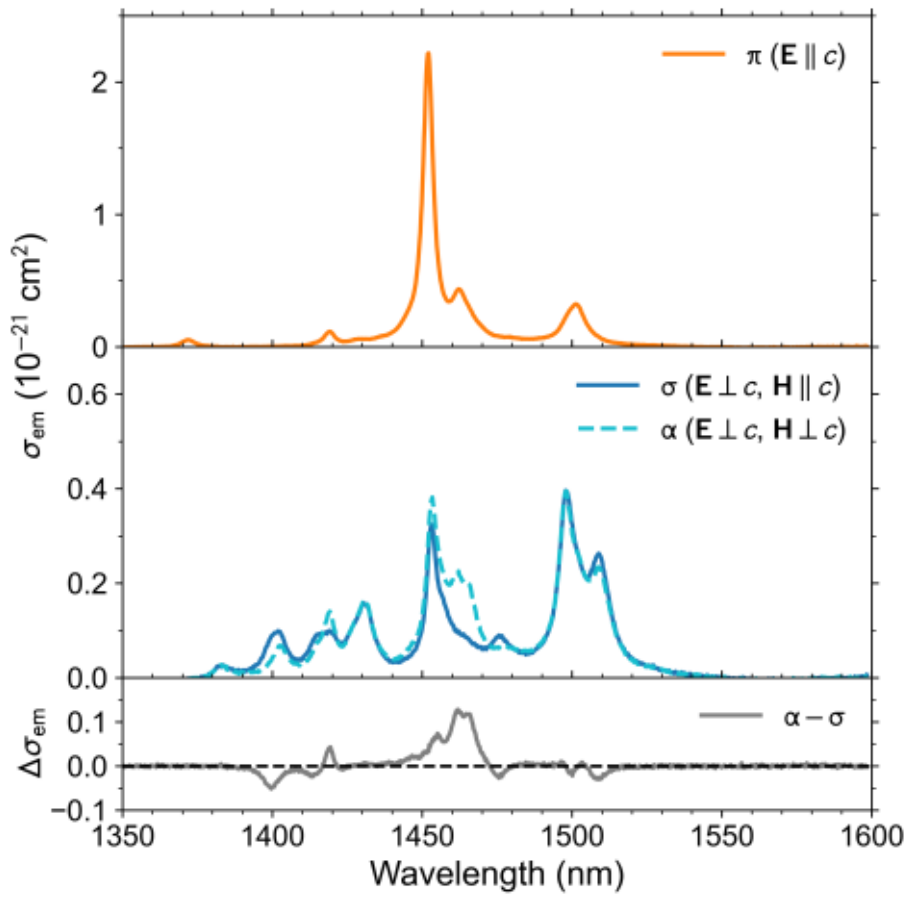

Fig. 5. Polarized emission cross-section spectra of Tm:YLF for the $^3H_4 \rightarrow {}^3F_4$ transition and corresponding differential cross-sections $\Delta\sigma$ between the σ- and α-polarization.

Figure 6 shows the emission cross-section spectra for the $^3H_4 \rightarrow {}^3H_5$ transition. The α-polarized spectrum shows higher noise, which arises from lower absorbed excitation power, since the stronger π-polarized absorption is not accessible in the *c*-cut sample required to measure the α-polarized emission. Thus, as-measured and smoothed data are shown in Fig. 6. The differential spectrum is calculated using the smoothed α-polarized spectrum. The difference between the σ- and α-polarization reaches $0.7\times10^{-21}$ cm$^2$ at 2310 nm. The spectral anisotropy factor $\kappa_{MD}$ is 26%.

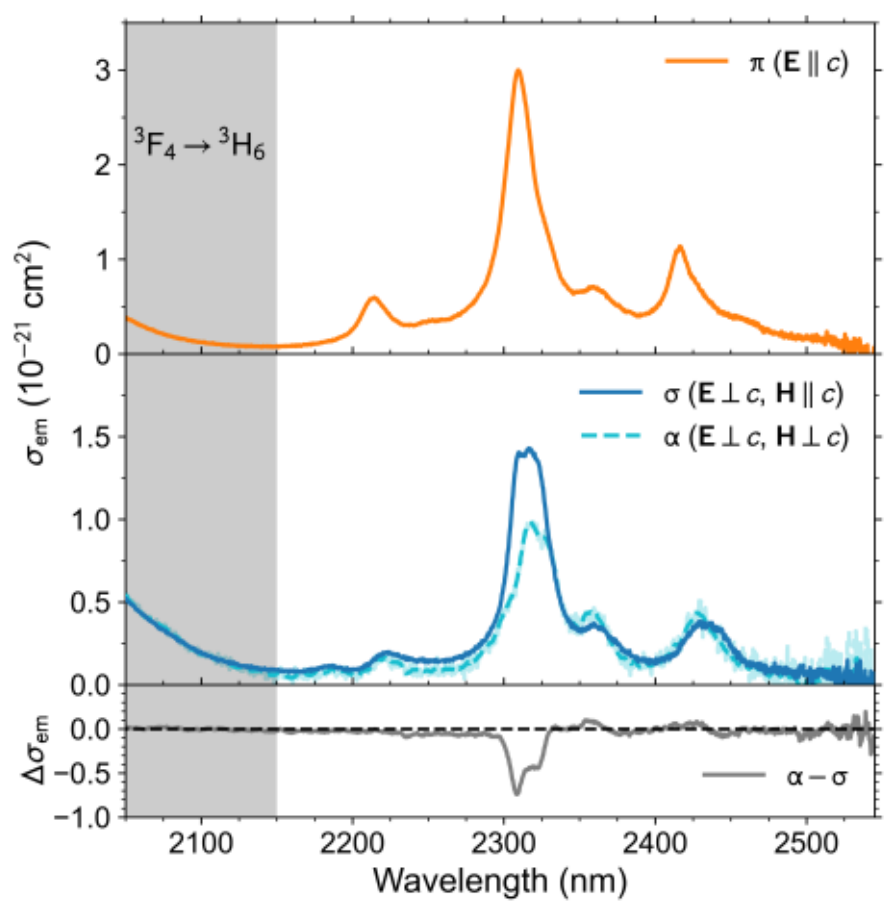

Fig. 6. Polarized emission cross-section spectra of Tm:YLF for the $^3H_4 \rightarrow {}^3H_5$ transition and corresponding differential emission cross-sections $\Delta\sigma_{em}$ between the σ- and α-polarization. Emission for wavelengths below 2150 nm, highlighted in grey, originates from the $^3F_4 \rightarrow {}^3H_6$ transition. For α-polarization, as-measured and smoothed spectra are shown.

### *4.3 Er:YLF*

Figures 7(a) and 7(b) show the absorption and emission cross-section spectra of Er:YLF for the $^4I_{15/2} \leftrightarrow {}^4I_{13/2}$ transition around 1.5 µm, respectively. The differential cross-section spectra $\Delta\sigma_{abs}$ and $\Delta\sigma_{em}$ exhibit similar trends, consistent with the reciprocity relation (cf. Eq. (5)). Differences up to $0.2\times10^{-20}$ cm$^2$ are observed between the σ- and α-polarized absorption and emission spectra. The spectral anisotropy factor is 5% for absorption and 12% for emission.

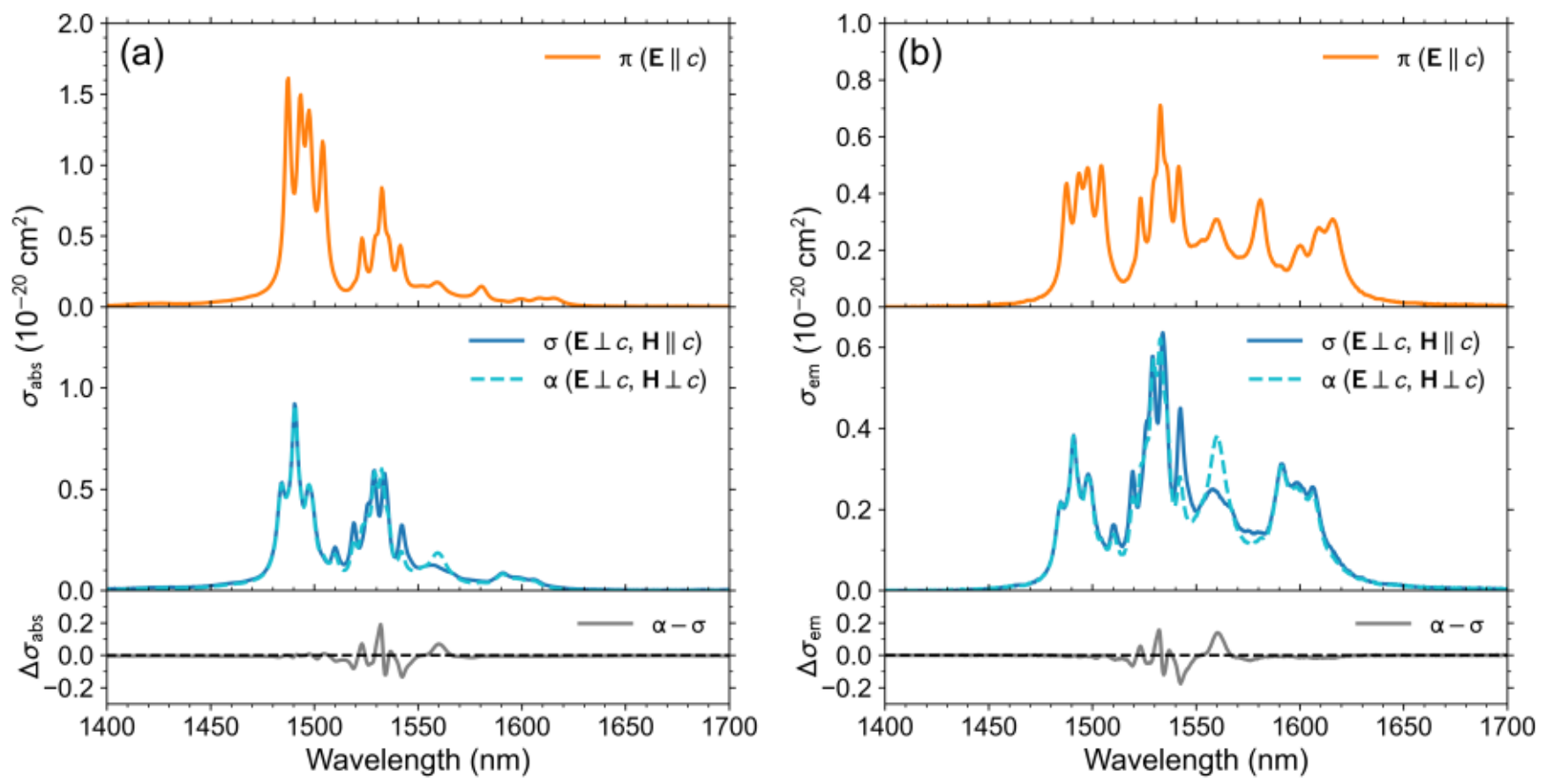

Fig. 7. Polarized absorption (a) and emission (b) cross-section spectra of Er:YLF for $^4I_{15/2} \leftrightarrow {}^4I_{13/2}$ transition and corresponding differential cross-sections $\Delta\sigma$ between the σ- and α-polarization.

The emission cross-section spectra for the $^4I_{11/2} \rightarrow {}^4I_{13/2}$ transition around 2.7 µm in Fig. 8 show differences between σ- and α-polarizations over the whole wavelength range. We observed differences up to $1\times10^{-20}$ cm$^2$. The spectral anisotropy factor is 23%. Note that the absolute emission cross-section values require further investigation, owing to the uncertainties in the radiative lifetime and branching ratio, which arise from the unknown non-radiative relaxation rate.

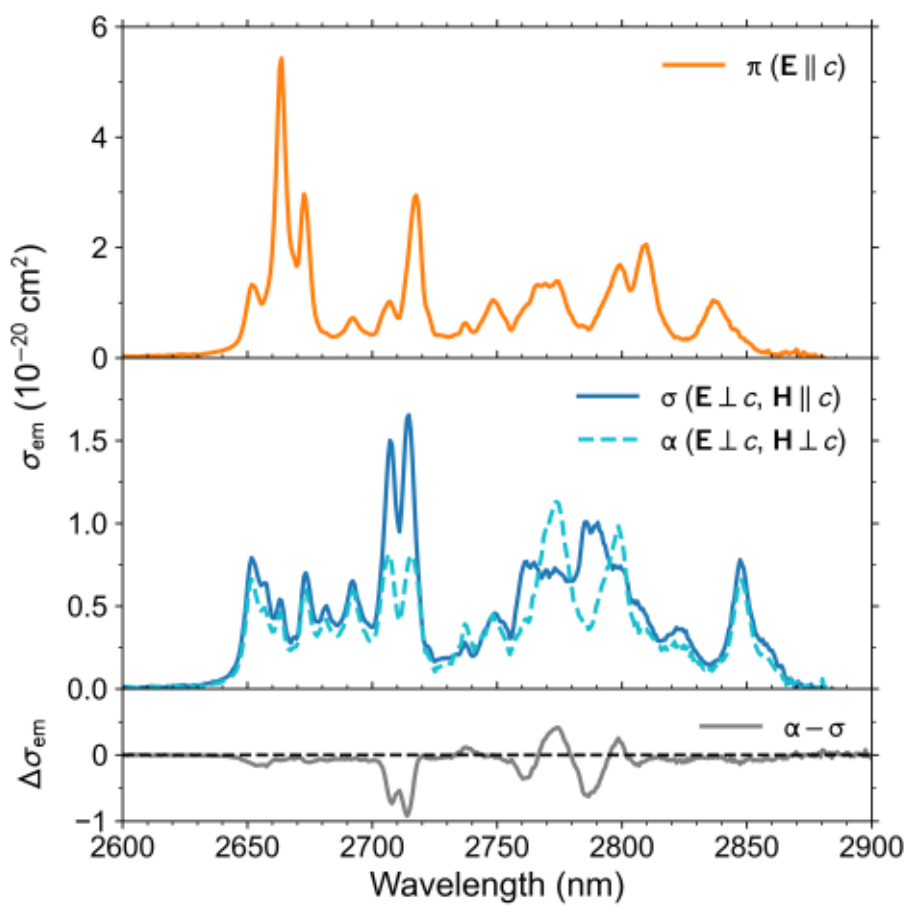

Fig. 8. Polarized emission cross-section spectra of Er:YLF for the $^4I_{11/2} \rightarrow {}^4I_{13/2}$ emission transition and corresponding differential emission cross-sections $\Delta\sigma_{em}$ between the σ- and α-polarization.

### *4.4 Ho:YLF*

Figure 9 shows the absorption and emission cross-section spectra of Ho:YLF for the $^5I_7 \leftrightarrow {}^5I_8$ transitions around 2.0 µm. Both absorption and emission are stronger in α- than in σ-polarization. Absorption and emission show differences in cross sections up to $0.35\times10^{-20}$ cm$^2$ and $0.55\times10^{-20}$ cm$^2$ at 1945 nm. The spectral anisotropy factors $\kappa_{MD}$ are 20% and 16% for absorption and emission, respectively.

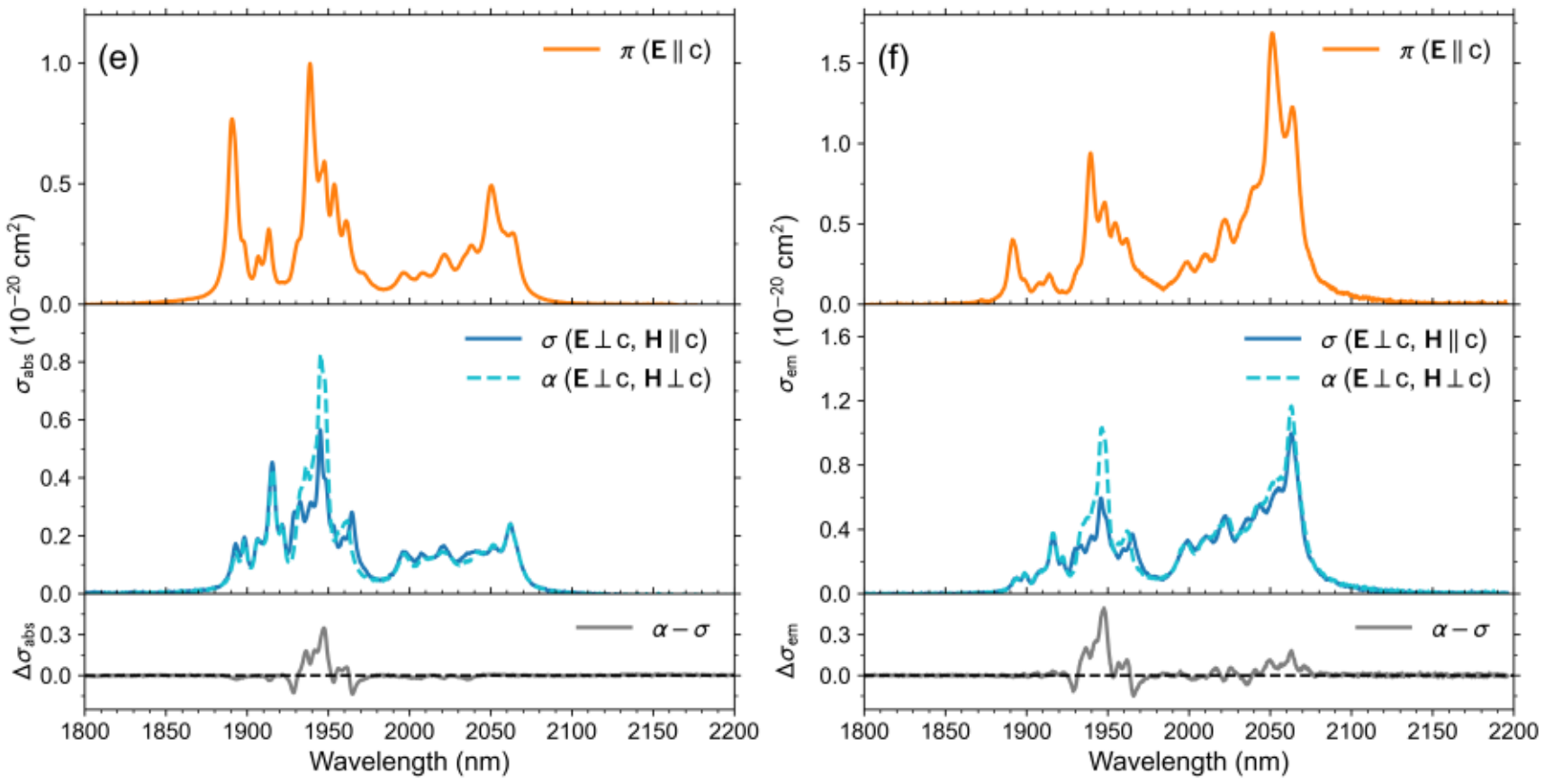

Fig. 9. Polarized absorption (a) and emission (b) cross-section spectra of Ho:YLF for the $^5I_7 \leftrightarrow {}^5I_8$ transition and corresponding differential cross-sections Δσ between the σ- and α-polarization.

Figure 10 shows the emission cross-section spectra for the $^5I_6 \rightarrow {}^5I_7$ transition. We applied smoothing to the α-polarized spectrum for the range >2900 nm. Differences in emission cross-section up to $2.4\times10^{-20}$ cm$^2$ at 2950 nm are observed between the σ- and α-polarized spectra. The spectral anisotropy factor $\kappa_{MD}$ is 76%, the highest among the transitions investigated in this work. Similar to the 2.7-µm transition of $Er^{3+}$, the absolute values of the emission cross-sections for this transition may need to be revisited.

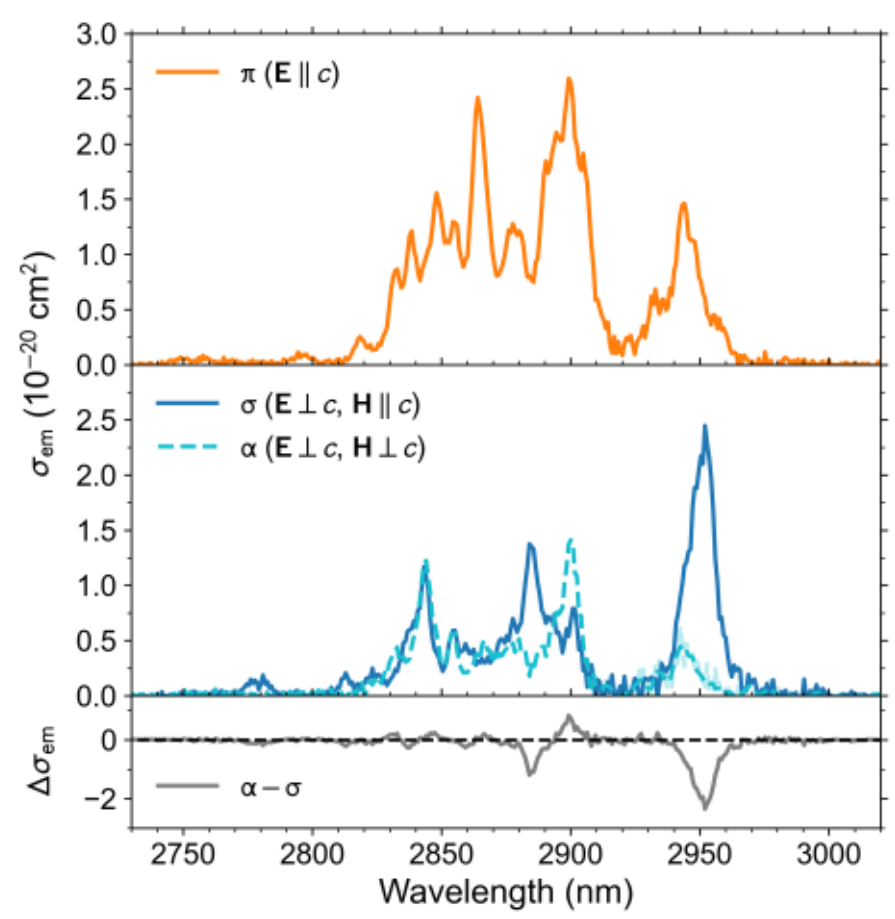

Fig. 10. Polarized emission cross-section spectra of Ho:YLF for the $^5I_6 \rightarrow {}^5I_7$ transition and corresponding differential emission cross-sections $\Delta\sigma_{em}$ between the σ- and α-polarization. For α-polarization, as-measured and smoothed (>2900 nm, light blue) spectra are shown.

## 5. Discussion

### 5.1 Magnetic-dipole branching ratio, selection rules and spectral anisotropy

Table 6 summarizes the relevant theoretical and experimental parameters obtained in this work. The value $\Delta\sigma_{em,max}/\sigma_{em,max}$ is hereby defined as the maximum difference in cross-sections between σ- and α-polarizations normalized to the highest peak cross-section among the three polarizations. We find that a high MD branching ratio does not necessarily lead to strong MD-induced spectral anisotropy, and vice versa.

The observed MD-induced spectral anisotropy is expected to originate from magnetic-polarization-dependent transition probabilities between Stark sub-levels within the manifolds, governed by the MD selection rules (see Tables 4 and 5). For the investigated transitions in $Yb^{3+}$ and $Er^{3+}$, half of the inter-Stark-level transitions are potentially magnetic-polarization dependent (Table 9 in Appendix 1). In contrast, for those in $Tm^{3+}$ and $Ho^{3+}$, this fraction is 75–78% (Table 9 in Appendix 1). However, we do not observe more significant MD-induced spectral anisotropy in $Tm^{3+}$ and $Ho^{3+}$ than in $Yb^{3+}$ and $Er^{3+}$. Furthermore, according to the selection rules for half-integer $J$ (Table 4), there is no inter-Stark-level transition that is exclusively allowed for magnetic polarization $\mathbf{H} \parallel c$ (σ-polarization). However, the magnetic-polarization-dependent spectra of $Yb^{3+}$ and $Er^{3+}$ show some peaks more prominent in $\mathbf{H} \parallel c$ (σ-polarization) than in $\mathbf{H} \perp$ c (α-polarization). These observations suggest that MD-induced spectral anisotropy does not originate solely from the selection rules. Further investigations, *e.g.*, low-temperature magnetic-polarization-dependent spectroscopy, are required to clarify other sources of MD-induced spectral anisotropy.

**Table 6. Calculated MD branching ratio $\beta_{MD}$, anisotropy factor $\kappa_{MD}$, maximum relative difference in cross-sections between σ- and α-polarizations, and error $\varepsilon$ defined by Eq. (9) for the seven investigated emission transitions of $Yb^{3+}$, $Tm^{3+}$, $Er^{3+}$, and $Ho^{3+}$ in YLF.**

| Ion | Transition | $\beta_{MD}$ (%) | $\kappa_{MD}$ (%) | $\Delta\sigma_{em,max}$ / $\sigma_{em,max}$ (%) | $\varepsilon$ (%) |
|---|---|---|---|---|---|
| $Yb^{3+}$ | $^2F_{5/2} \rightarrow {}^2F_{7/2}$ | 11 | 7 | 11 | -1 |
| $Tm^{3+}$ | $^3H_4 \rightarrow {}^3F_4$ | 67 | 21 | 6 | -2 |
| | $^3H_4 \rightarrow {}^3H_5$ | 46 | 23 | 26 | 2 |
| $Er^{3+}$ | $^4I_{13/2} \rightarrow {}^4I_{15/2}$ | 30 | 12 | 25 | 1 |
| | $^4I_{11/2} \rightarrow {}^4I_{13/2}$ | 51 | 23 | 17 | 5 |

| Ho$^{3+}$ | $^5I_7 \rightarrow {}^5I_8$ | 72 | 16 | 29 | -4 |
|---|---|---|---|---|---|
| | $^5I_6 \rightarrow {}^5I_7$ | 88 | 76 | 91 | 13 |

### *5.2 Impact of MD contributions on calculated transition cross-sections*

The impact of the MD-induced spectral anisotropy on the calculation of the π- and σ-polarized emission cross-sections based on the F–L formula can be quantified by:

$$\varepsilon = \frac{\int \lambda(I_\pi + I_\sigma + I_\alpha)d\lambda}{\int \lambda(I_\pi + 2I_\sigma)d\lambda} - 1 = \frac{\int \lambda(I_\alpha - I_\sigma)d\lambda}{\int \lambda(I_\pi + 2I_\sigma)d\lambda}. \quad (9)$$

The quantity $\varepsilon$ represents the error in the emission cross-section of uniaxial crystals by only considering π- and σ-polarizations. A positive value of $\varepsilon$ signifies that neglecting α-polarization overestimates the emission cross-section and vice versa. Table 6 summarizes the calculated error $\varepsilon$ for the seven emission transitions, showing that the assumption of $I_\sigma(\lambda) \approx I_\alpha(\lambda)$ is valid in most cases, as $\varepsilon$ is within ±5%, even for these transitions selected for their expected significant MD contributions. An exception is the $^5I_6 \rightarrow {}^5I_7$ transition of Ho:YLF at 2.9 µm, where $\varepsilon$ is 13%. However, note that the MD-induced spectral anisotropy can still be significant despite $\varepsilon$ being small. For instance, despite $\varepsilon$ of only 1%, the $^4I_{13/2} \rightarrow {}^4I_{15/2}$ emission transition in Er:YLF exhibits a difference in the emission cross section between the σ- and α-polarizations as high as 25% of the highest peak cross section at ≈1560 nm.

The Judd–Ofelt analysis uses the integrated absorption cross-section spectra $\int \sigma_{\text{abs}}(\lambda)d\lambda$ for each transition. To our knowledge, MD-induced spectral anisotropy has not been considered in such analyses. Our results suggest that, for accurate Judd–Ofelt analyses of uniaxial crystals with significant MD-induced anisotropy, three polarization-dependent spectra should be used; for biaxial crystals, even six spectra may be required.

### *5.3 Other host crystals*

Here, we discuss the consequences of our findings for other host materials, including biaxial crystals, for Yb$^{3+}$ as the active ion. As mentioned in Section 2, the MD oscillator strength and spontaneous emission rate scale with $n^3$, and a long radiative lifetime indicates weak ED transitions, increasing MD contributions. Table 7 lists some relevant parameters for various uniaxial and biaxial host crystals. This table extends Table IV in [8]. It shows that Yb$^{3+}$-doped fluoride crystals typically exhibit longer radiative lifetimes, while their refractive indices are lower than those of oxides. Among the crystals listed in Table 7, Yb-doped $YCa_4O(BO_3)_3$ (Yb:YCOB) shows the largest calculated MD branching ratio $\beta_{\text{MD}}$ of 55%. The second largest value is found for Yb-doped $LaLuO_3$, for which we recently identified strong MD-induced spectral anisotropy [16]. Similar to Eq. (6), we define the MD-induced spectral anisotropy factor $\kappa_{\text{MD}}$ for biaxial crystals as follows:

$$\kappa_{\text{MD}} = \frac{2\int |\sigma_{\text{X,Y}}(\lambda) - \sigma_{\text{X,Z}}(\lambda)| + |\sigma_{\text{Y,X}}(\lambda) - \sigma_{\text{Y,Z}}(\lambda)| + |\sigma_{\text{Z,X}}(\lambda) - \sigma_{\text{Z,Y}}(\lambda)| d\lambda}{\sum_{i,j} \int \sigma_{i,j}(\lambda) d\lambda}, \quad (7)$$

where $\sigma_{i,j}$ corresponds to a cross-section spectrum for **E** ∥ $i$ and **H** ∥ $j$ ($i$ and $j$ are X, Y, or Z, $i \neq j$). Using this formula, $\kappa_{\text{MD}}$ of Yb:$LaLuO_3$ is calculated to be ≈6%.

**Table 7. Polarization-averaged refractive index $\bar{n}$ at a wavelength of ≈1 µm, radiative lifetime $\tau_{\text{rad}}$, and MD branching ratio $\beta_{\text{MD}}$ calculated for Yb$^{3+}$ in various anisotropic host crystals.**

| Host | Optical class | $\bar{n}$ | $\tau_{\text{rad}}$ (ms) | $\beta_{\text{MD}}$ (%) | Ref. |
|---|---|---|---|---|---|
| $LiYF_4$ (YLF) | uniaxial | 1.44 | 2.2 | 11 | [27,28] |
| $BaY_2F_8$ | biaxial | 1.49 | 2.04 | 11 | [38,39] |
| $LaF_3$ | uniaxial | 1.6 | 2.22 | 15 | [38,40] |
| $YVO_4$ | uniaxial | 2.03 | 0.247 | 3 | [41,42] |
| $GdVO_4$ | uniaxial | 2.1 | 0.345 | 5 | [42,43] |
| $LuVO_4$ | uniaxial | 2.24 | 0.256 | 5 | [42,44] |
| $YAlO_3$ (YAP) | biaxial | 1.93 | 0.72 | 9 | [38,45] |
| $Y_2SiO_5$ (YSO) | biaxial | 1.79 | 1.04 | 10 | [38,46] |
| $LaLuO_3$ | biaxial | ≈1.95 | 1.46 | 18 | [16] |
| $CaGdAlO_4$ (CALGO) | uniaxial | 1.92 | 0.44 | 5 | [47,48] |
| $YCa_4O(BO_3)_3$ (YCOB) | biaxial | ≈1.7 | 2.2 | 55 | [49,50] |
| $KY(WO_4)_2$ (KYW) | biaxial | 2.0 | 0.6 | 8 | [51,52] |
| $KGd(WO_4)_2$ (KGW) | biaxial | 2.0 | 0.6 | 8 | [51,52] |
| $LuPO_4$ | uniaxial | 1.9 | 0.83 | 9 | [38,53] |

To experimentally verify the MD-induced spectral anisotropy in biaxial crystals, we measured all six polarized absorption spectra for the fluoride crystal Yb:$BaY_2F_8$ and the oxide crystal Yb:$Y_2SiO_5$, as shown in Fig. 11. According to Table 7, their MD branching ratios are comparable; however, the MD-induced spectral anisotropy factor $\kappa_{\mathrm{MD}}$ is ≈5% for Yb:$BaY_2F_8$, whereas it is negligibly small for Yb:$Y_2SiO_5$, although a small difference may be present at 990 nm. These results demonstrate that the degree of MD-induced anisotropy depends heavily on the host crystal, even for comparable calculated MD branching ratios.

It should be noted that $Yb^{3+}$ shows stronger electron-phonon coupling [54], resulting in broader spectra compared with other trivalent lanthanide ions. We speculate that vibronic transitions may obscure the MD-induced spectral anisotropy based on resonant inter-Stark-level transitions. The MD-induced spectral anisotropy might be more pronounced at low temperatures, and thus, low-temperature spectra of Yb:$Y_2SiO_5$ may show a stronger effect. Temperature-dependent measurements of the MD-induced anisotropy are therefore required to clarify the roles of line broadening and phonon coupling.

As observed in Yb:$LaLuO_3$ [16], the impact of MD transitions can be pronounced if a host crystal possesses inversion-symmetry sites at which ED transitions are forbidden. Therefore, in such a host crystal, MD-induced spectral anisotropy may still be prominent despite small calculated MD branching ratios.

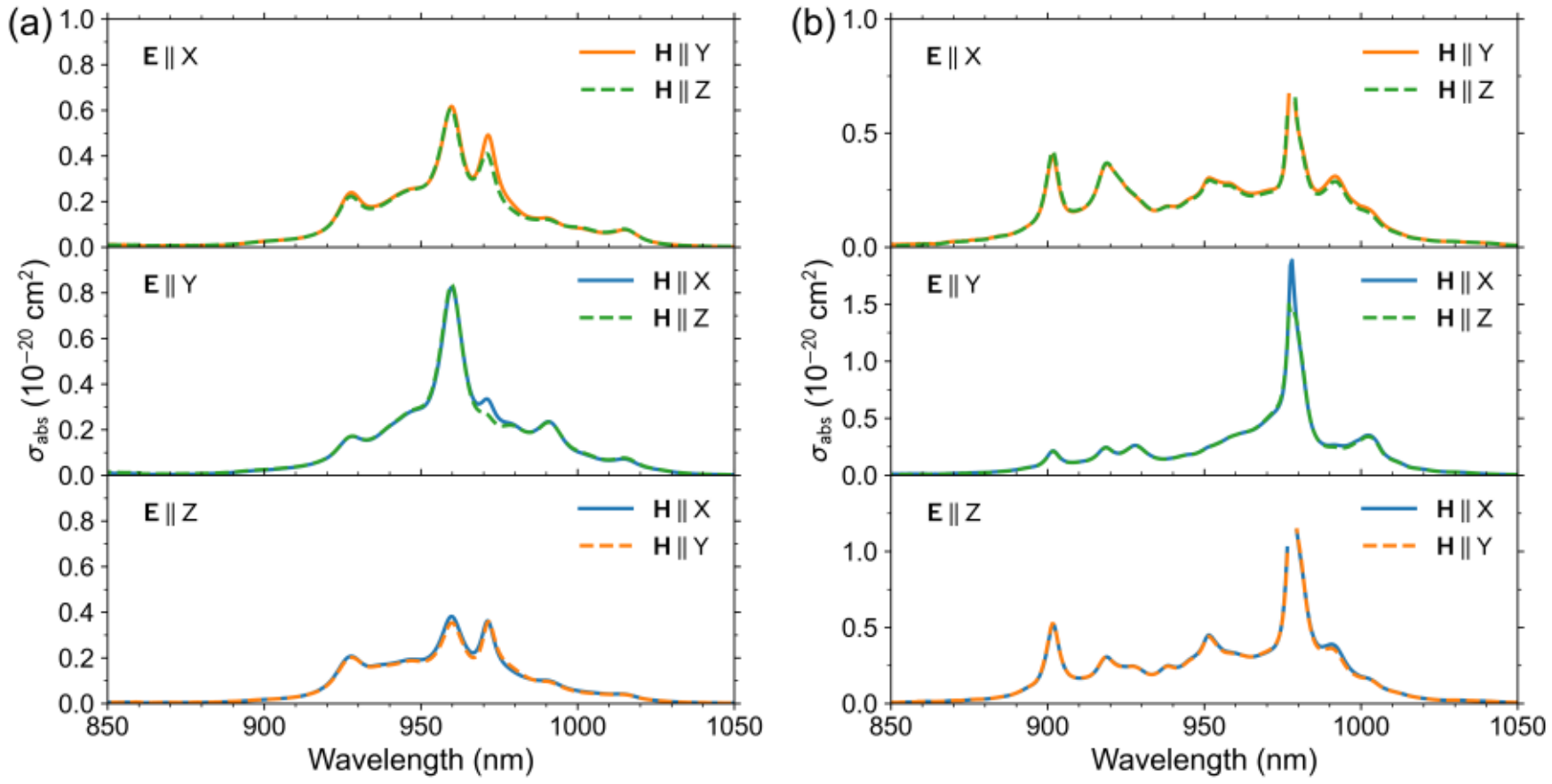

Fig. 11. Polarized absorption cross-section spectra of $Yb^{3+}$-doped $BaY_2F_8$ (a) and $Y_2SiO_5$ (b) crystals for the transition ${}^2F_{7/2} \rightarrow {}^2F_{5/2}$. For Yb:$Y_2SiO_5$, absorption cross-sections around 980 nm are not shown except **E** || Y, **H** || X, because the transmittance was too low for reliable calculation of the cross-sections owing to the thick samples.

### *5.4 Implications for solid-state lasers*

The polarization dependence of the absorption, emission, as well as gain cross-sections has important implications for solid-state lasers. Although the highest emission cross-sections are often observed for π-polarization in YLF, as seen in Figs. 3–10, laser operation based on the electric polarization **E** ⊥ *c* can offer the advantage of weakly positive thermal lensing [55]. From this perspective, the use of a *c*-cut YLF laser crystal is attractive. In such a geometry, accurate knowledge of the transition cross-section spectra for α-polarization becomes essential. However, the differences have rarely been considered for the design of solid-state lasers so far, although our results show that this distinction can be significant for some transitions. For example, in Tm:YLF, the σ- and α-polarized peak emission cross-sections are nearly identical in the wavelength range relevant for laser operation between 1498 nm and 1509 nm [25]. In this case, knowing the σ-polarized spectrum is sufficient. In contrast, the 2.9-µm transition in Ho:YLF shown in Fig. 10 shows more than 90% anisotropy between the α- and σ-polarized peak emission cross-sections near 2950 nm, which is essential to consider for the design of lasers based on the corresponding transition.

For biaxial systems, the orientation of laser crystals is usually selected to access the polarization maximizing the pump absorption efficiency and the optical gain; however, this selection is based solely on the three previously considered electric-polarization-dependent spectra. Given a potential MD-induced spectral anisotropy, the choice among the six instead of the three possible polarizations may need reconsideration.

## 6. Conclusion

We have demonstrated that MD contributions can cause a pronounced and previously underappreciated spectral anisotropy in lanthanide-doped anisotropic crystals. Magnetic-polarization-dependent spectroscopy revealed the MD-induced spectral anisotropy in trivalent $Yb^{3+}$, $Tm^{3+}$, $Er^{3+}$, and $Ho^{3+}$ ions in the uniaxial crystal YLF. This showed the need to consider α-polarization (**E** ⊥ *c*, **H** ⊥ *c*), in addition to the π- (**E** ∥ *c*, **H** ⊥ *c*) and σ-polarizations (**E** ⊥ *c*, **H** ∥ *c*) for a complete spectroscopic characterization of these anisotropic crystals for a wide range of transitions in lanthanide doping ions.

Although theoretically calculated MD oscillator strengths and spontaneous emission rates are useful for identifying transitions that may exhibit MD-induced spectral anisotropy, our experimental results show that the resulting anisotropy cannot be predicted from these calculations; thus, experimental characterization is essential. We also found that MD selection rules for inter-Stark-level transitions do not explain the experimental observations well. This implies that other factors contribute to the resulting magnetic-polarization dependence, requiring further investigation.

Considering magnetic-polarization-dependent spectra in the F–L formula enables a refinement of previously reported emission cross-section values, which mostly ignored the MD-induced anisotropy. We find that the resulting error in the calculated emission cross section due to the ignored α-polarized spectrum, $\varepsilon$, remains below ±5% except for the 2.9-µm emission of $Ho^{3+}$, which shows an error of 13%. On the other hand, the peak cross-sections can differ even for transitions with small $\varepsilon$. The cross-section difference between σ- and α-polarizations normalized by the maximum peak cross section reaches 91% for the 2.9-µm emission of $Ho^{3+}$. The latter may have significant implications for the modeling of solid-state lasers based on the corresponding transitions.

The magnetic-polarization dependence is also likely present in other anisotropic crystals, particularly those with weak ED transitions, typically found in low-crystal-field hosts, particularly with high site symmetry. This may require the re-evaluation of polarization-dependent spectra in various laser hosts.

## Appendix 1. Selection rules for inter-Stark-level magnetic-dipole transitions under $S_4$ point-group symmetry

We tabulate degeneracies given by $2J+1$, numbers of split Stark levels under $S_4$ point-group symmetry, and corresponding irreducible representations in Koster notation for the manifolds relevant to this work, based on full-rotation-group compatibility under this point group [33]. Using Table 8 and the selection rules in Tables 4 and 5, we further tabulate, for the investigated inter-manifold transitions, the total numbers of inter-Stark-level transitions and the numbers of transitions exclusively allowed for one magnetic polarization, $\mathbf{H} \parallel c$ or $\mathbf{H} \perp c$, which potentially cause MD-induced spectral anisotropy.

**Table 8. Degeneracies, number of split Stark levels under $S_4$ point-group symmetry, and corresponding irreducible representations of manifolds of the $Yb^{3+}$, $Tm^{3+}$, $Er^{3+}$, and $Ho^{3+}$ manifolds relevant to this work.**

| Ion | Manifold *SLJ* | Degeneracy $2J+1$ | Number of split Stark levels | Representation |
|---|---|---|---|---|
| $Yb^{3+}$ | $^2F_{7/2}$ | 8 | 4 | $2\Gamma_{5,6} + 2\Gamma_{7,8}$ |
| | $^2F_{5/2}$ | 6 | 3 | $\Gamma_{5,6} + 2\Gamma_{7,8}$ |
| $Tm^{3+}$ | $^3H_6$ | 13 | 10 | $3\Gamma_1 + 4\Gamma_2 + 3\Gamma_{3,4}$ |
| | $^3F_4$ | 9 | 7 | $3\Gamma_1 + 2\Gamma_2 + 2\Gamma_{3,4}$ |
| | $^3H_5$ | 11 | 8 | $3\Gamma_1 + 2\Gamma_2 + 3\Gamma_{3,4}$ |
| | $^3H_4$ | 9 | 7 | $3\Gamma_1 + 2\Gamma_2 + 2\Gamma_{3,4}$ |
| $Er^{3+}$ | $^4I_{15/2}$ | 16 | 8 | $4\Gamma_{5,6} + 4\Gamma_{7,8}$ |
| | $^4I_{13/2}$ | 14 | 7 | $3\Gamma_{5,6} + 4\Gamma_{7,8}$ |
| | $^4I_{11/2}$ | 12 | 6 | $3\Gamma_{5,6} + 3\Gamma_{7,8}$ |
| $Ho^{3+}$ | $^5I_8$ | 17 | 12 | $4\Gamma_1 + 4\Gamma_2 + 4\Gamma_{3,4}$ |
| | $^5I_7$ | 15 | 11 | $3\Gamma_1 + 4\Gamma_2 + 4\Gamma_{3,4}$ |
| | $^5I_6$ | 13 | 10 | $3\Gamma_1 + 4\Gamma_2 + 3\Gamma_{3,4}$ |

**Table 9. Total numbers of inter-Stark-level transitions for the investigated inter-manifold transitions in $Yb^{3+}$, $Tm^{3+}$, $Er^{3+}$, and $Ho^{3+}$, together with numbers of inter-Stark-level transitions allowed only for one magnetic polarization, H || *c* or H $\perp$ *c*. These polarization-selective transitions are potential sources of magnetic-dipole-induced spectral anisotropy.**

| Ion | Transition | Total number of inter-Stark-level transitions | Number of transitions allowed only for $\mathbf{H} \parallel c$ (σ-polarization) | Number of transitions allowed only for $\mathbf{H} \perp c$ (π and α-polarizations) |
|---|---|---|---|---|
| $Yb^{3+}$ | $^2F_{7/2} \leftrightarrow {}^2F_{5/2}$ | 12 | 0 | 6 |
| $Tm^{3+}$ | $^3H_6 \rightarrow {}^3H_5$ | 80 | 26 | 36 |
| | $^3H_4 \rightarrow {}^3H_5$ | 56 | 19 | 25 |
| | $^3H_4 \rightarrow {}^3F_4$ | 49 | 17 | 20 |
| $Er^{3+}$ | $^4I_{15/2} \leftrightarrow {}^4I_{13/2}$ | 56 | 0 | 28 |
| | $^4I_{11/2} \rightarrow {}^4I_{13/2}$ | 42 | 0 | 21 |
| $Ho^{3+}$ | $^5I_8 \leftrightarrow {}^5I_7$ | 132 | 44 | 60 |
| | $^5I_6 \rightarrow {}^5I_7$ | 110 | 37 | 49 |

**Funding.**

Deutsche Forschungsgemeinschaft (520253663); Leibniz Association (J183/2023).

**Acknowledgment.**

The authors acknowledge the technical support of Celine Kapella and John Stahl in the growth, as well as Albert Kwasniewski (all from IKZ) for crystal orientation of the fluoride crystals used in this work using X-ray Laue diffraction.

**Disclosures.** The authors declare no conflicts of interest.

**Data availability.**

Data underlying the results presented in this paper are available from the authors upon request.